 \documentclass[final,5p,times,twocolumn]{elsarticle}

 \usepackage{graphics}
 \usepackage{graphicx}

\usepackage{amssymb}
\usepackage{amsmath}

\journal{Journal of Molecular Structure}

\begin{document}

\begin{frontmatter}



\title{Rotational spectroscopy of singly $^{13}$C substituted isotopomers of propyne 
       and determination of a semi-empirical equilibrium structure}


\author[Koeln]{Holger S.P.~M\"uller\corref{cor}}
\ead{hspm@ph1.uni-koeln.de}
\cortext[cor]{Corresponding author.}
\author[Koeln]{Sven Thorwirth}
\author[Koeln]{Frank Lewen}

\address[Koeln]{I.~Physikalisches Institut, Universit{\"a}t zu K{\"o}ln, 
  Z{\"u}lpicher Str. 77, 50937 K{\"o}ln, Germany}

\begin{abstract}

Submillimeter spectra of three isotopomers of propyne containing one $^{13}$C atom were 
recorded in natural isotopic composition in the region of 426~GHz to 785~GHz. Additional 
measurements were carried out near 110~GHz. Combining these with earlier data resulted 
in greatly improved spectroscopic parameters which permit reliable extrapolations up to 
about 1.5~THz. Coupled cluster quantum-chemical calculations were carried out in order to 
assess the differences between equilibrium and ground state rotational parameters of these 
and many other isotopic species to evaluate semi-empirical equilibrium structural parameters.
In addition, we estimated the main spectroscopic parameters of the isotopomers of propyne 
with two $^{13}$C atoms, which have not yet been studied in the laboratory, but which may 
be detectable in astronomical sources with a large amount of $^{13}$C compared to the dominant 
$^{12}$C.

\end{abstract}

\begin{keyword}  

rotational spectroscopy \sep 
interstellar molecule \sep
propyne \sep
symmetric top molecule \sep
quantum-chemical calculation \sep
structural parameters


\end{keyword}

\end{frontmatter}




\section{Introduction}
\label{introduction}

The detection of propyne, CH$_3$CCH, which is also known as methylacetylene, in the 
interstellar medium (ISM) was reported in 1973 \cite{MeCCH_det_SgrB2-T_1973}, making 
it one of the molecules discovered early by radio astronomical means. The observations 
were made toward Sagittarius (Sgr) B2, one of the most massive star-forming regions 
in our Galaxy and located close to its center \cite{SgrB2-properties_2016}. 
Even early observations of propyne, e.g. \cite{MeCCH_more-lines_sources_T_1981}, 
were employed to infer temperature and density in this source. 
Symmetric top rotor molecules, such as propyne, are particularly suited to derive 
temperature conditions in the ISM because transitions with the same $J$ and different 
$K$ quantum numbers occur in a narrow frequency range, but sample a large range of 
energies. Methyl cyanide, CH$_3$CN, is another example of a molecule frequently used 
for this purpose \cite{MeCN_det_T_1971,OMC-1-T_MeCN_1981}. Temperatures of other 
star-forming regions were determined via early propyne observations, such as 
Orion and DR~21 \cite{T-kin_MeCCH-vs-MeCN_1983}, and of many other sources, 
including the cold and dense star-less core TMC-1 
\cite{T-kin_div-hot-core_TMC-1_1984a,T-kin_div-hot-core_TMC-1_1984b}. The molecule was 
also detected in near-by galaxies, such as M82 and NGC253 \cite{propyne_extragal_1991}, 
in more distant galaxies, such as the foreground galaxy at $z \approx 0.89$ in the 
direction of the quasar PKS 1810$-$211 \cite{div_z_0p89_2011}, in translucent molecular 
clouds \cite{propyne_etc_translucent_2000}, and in the envelopes of late-type stars, 
such as the protoplanetary nebular CRL618 \cite{CRL618_abundances_2007}, the asymptotic 
giant branch star CW~Leonis \cite{propyne_etc_CW-Leo_2008} and the planetary nebula 
K4$-$47 \cite{PN_K4-47_2019}. Infrared observations with Voyager led to the detection 
of CH$_3$CCH in the atmosphere of Titan \cite{propyne_propane_Titan_1981}; the ISO 
satellite was employed later to discover propyne in the atmospheres of Saturn 
\cite{propyne_etc_Saturn_1997}, Jupiter \cite{propyne_etc_Jupiter_2000}, and Uranus 
\cite{propyne_etc_Uranus_2006}. More recently, propyne was also observed in the atmosphere 
of Titan with the Atacama Large Millimeter/submillimeter Array (ALMA) 
\cite{propyne_Titan_w-ALMA_2015}.

Propyne is so abundant in some astronomical sources that minor isotopic species 
were discovered as well. Even though CH$_3$C$^{13}$CH was found serendipitously 
\cite{also_propyne_13C-1_1987}, all isotopomers containing one $^{13}$C were observed 
later \cite{PN_K4-47_2019,12C-13C_SgrB2N_2017,L483_3mm_2019}. The deuterated isotopologs 
CH$_2$DCCH \cite{CH2DCCH_det_1992} and CH$_3$CCD \cite{CH3CCD_det_2005} were also 
detected. There is even evidence that $^{13}$CH$_2$DCCH was seen \cite{L483_3mm_2019}.

Observations of molecules containing $^{13}$C are important diagnostic tools because 
the $^{12}$C/$^{13}$C ratio in space differs considerably from the terrestrial value 
of 89 \cite{isot-comp_2009}. It is as low as about 20 to 25 in the Galactic center region 
\cite{12C-13C_SgrB2N_2017,13C-VyCN_2008,EMoCA_2016,RSH_ROH_2016}, increases to about 
68 in the Solar neighborhood and even further in the outskirts of the Milky Way 
\cite{12C-13C_SgrB2N_2017,galactic_isotopic-ratios_1994,12C-13C-gradient_2005}. 
Much lower $^{12}$C/$^{13}$C ratios than 20 were found in the envelopes of some late-type 
stars, for example $\sim$10 for CRL618 \cite{CRL618_abundances_2007,CRL618_12C-13C_etc_2003}, 
$\sim$4 for CK~Vulp{\'e}cula \cite{Nova-Vul_1670_12C-13C_etc_2015}, and $\sim$2 for 
K4$-$47 \cite{PN_K4-47_2019,PN_K4-47_12C-13C_etc_2018}.

A recent observational study of the protostellar binary IRAS 16293$-$2422 with ALMA 
\cite{propyne_IRAS-16293_2019} found only evidence for propyne in the hot corinos, the warmer 
and denser parts of the molecular clouds surrounding the protostars. The rotational 
temperatures of $\sim$100~K appear to be typical for propyne and several other molecules 
in such environments. Comparison with single dish observations, however, suggest that 
about half of the propyne must occur on larger scales in these molecular clouds such that 
interferometric observations (here with ALMA) filter out this part of propyne. 
Concurrent astrochemical modeling results suggest that both grain as well as gas phase 
chemistry needs to be invoked to explain the levels of propyne observed in that study 
\cite{propyne_IRAS-16293_2019}.

It is necessary to know rest frequencies of a certain molecule with sufficient accuracy 
to be able to identify it in space. Such rest frequencies are usually based on laboratory 
spectroscopic investigations and are frequently provided in databases such as the 
Cologne Database for Molecular Spectroscopy, CDMS\footnote{See https://cdms.astro.uni-koeln.de/.} 
\cite{CDMS_2016}. Recent investigations of $^{13}$C-containing isotopologs of molecules detected 
in space include for example dimethyl ether with one and two $^{13}$C \cite{13C-DME_rot_2013}, 
the acetaldehyde isotopomers with one $^{13}$C \cite{13C-MeCHO_rot_2015}, $^{13}$CH$_3$NH$_2$ 
\cite{13C-MeNH2_rot_2016}, isotopic acetone with emphasis on CH$_3$$^{13}$C(O)CH$_3$ 
\cite{13C-acetone_2019}, isotopic $c$-C$_3$H$_2$ with several doubly substituted species 
\cite{c-C3H2-isos_rot_2012}, and isotopic methyl cyanide in the $\varv_8 = 1$ excited state 
\cite{MeCN_isos_v8_rot_2016}.

The rotational spectrum of propyne was reported as early as 1950. Trambarulo and Gordy 
recorded the spectra of several isotopic species up to 52~GHz and determined a ground state 
effective ($r_0$) structure \cite{propyne_div-ios_rot_1950}. Thomas et al. revisited the 
$r_0$ structure on the basis of additional isotopologs and additional measurements for some 
of the earlier studied isotopologs \cite{MeCCH_MeCN_div-isos_rot_1955}. 
The latest investigation into the experimental structural parameters of propyne was 
carried out by Le Guennec et al. \cite{CH2DCCH_CH3CCD_isos_rot_1993}. They employed 
isotopically enriched samples of CH$_2$DCCH and CH$_3$CCD to record not only extensive 
spectra of these two isotopic variants, but also of the respective isotopomers with one 
$^{13}$C between 144 and 471~GHz.

Dubrulle et al. extended the measurements of CH$_3$CCH and its isotopomers with one 
$^{13}$C to 240~GHz \cite{propyne_wo_w-13C_rot_1978}. Further investigations of the main 
isotopic species reached quantum numbers of $K = 21$ and $J' = 70$ at frequencies up to 
1.2~THz \cite{CH3CCH_rot_2000,CH3CCH_gs_vib_rot_IR_2002,MeC2_4H_rot_2008}. Furthermore, 
the rotational spectroscopy of CH$_3$CCH in low-lying excited vibrational states was 
decisive to untangle rotation-vibration interactions between different polyads of 
propyne, e.g., between the lowest excited state $\varv _{10} = 1$ on one hand and 
$\varv _{10} = 2$ and $\varv _{9} = 1$ on the other 
\cite{CH3CCH_gs_vib_rot_IR_2002,CH3CCH_v39_2004,MeCCH_10mue_2009}. 
Similar interactions were analyzed for CH$_3$NC \cite{MeNC_v8le2_2011} and for 
CH$_3$CN \cite{MeCN_v8le2_2015}. 
The dipole moments \cite{MeCCH-D0-4_dip_1966,CD3CCH_etc_rot_dip_1970}, molecular $g$ 
values \cite{MeCCH-isos_Zeeman_1969}, hyperfine structure parameters 
\cite{CH3CCD_CD3CCH_rot_HFS_1982} and other properties of differently deuterated isotopic 
species of propyne were also studied.

The strongest transitions of propyne at $T = 100$~K occur around 325~GHz, but the lines are 
still quite strong above 500~GHz. The transition frequencies of propyne isotopomers with 
one $^{13}$C become increasingly uncertain at such frequencies because earlier measurements 
only reached $\sim$240~GHz \cite{propyne_div-ios_rot_1950,propyne_wo_w-13C_rot_1978}. 
Therefore, we have extended the measurements of these isotopic species up to 785~GHz. 
The resulting spectroscopic parameters permit reasonable extrapolation to about 1.5~THz, 
which is probably beyond the needs of radio astronomers.
In addition, we carried out quantum-chemical calculations to determine semi-empirical 
structural parameters and spectroscopic parameters of propyne isotopomers with two $^{13}$C, 
which have not yet been studied in the laboratory.


\section{Experimental details}
\label{exptl_details}

The rotational spectra were recorded at room temperature in static mode employing Pyrex 
glass cells $\sim$4~m in length and with an inner diameter of 100~mm. A commercial sample 
of propyne was used in natural isotopic composition and at pressures of mostly 1~Pa, raised 
up to 4~Pa for weaker transitions in the submillimeter region. The window material was Teflon 
at lower frequencies, whereas high-density polyethylene was used at higher frequencies.
Frequency modulation was used throughout with demodulation at $2f$, causing an isolated 
line to appear close to a second derivative of a Gaussian.

Measurements between 426 and 785~GHz were carried out with the Cologne Terahertz 
Spectrometer \cite{THz-BWO_1994} employing three different phase-locked backward 
wave oscillators (BWOs) as sources and a liquid helium cooled InSb hot-electron 
bolometer (QMC) as detector. Accuracies of 10~kHz and better can be achieved for 
isolated lines with good signal-to-noise ratio and very symmetric line shape, 
e.g., \cite{SiS_rot_2007}. Additional recordings near 100~GHz were performed with 
a BWO based 3~mm synthesizer AM-MSP~2 (Analytik \& Me{\ss}technik GmbH, Chemnitz, 
Germany) as source and a Schottky-diode as detector. Uncertainties as low as 1~kHz 
can be reached with this setup under favorable conditions \cite{HC3N_rot_IR_2017}. 
Only indivial lines were recorded in most instances.

\section{Quantum-chemical calculations}
\label{qcc}

Coupled cluster calculations with singles and doubles excitations augmented by a perturbative 
correction for triple excitations, CCSD(T) \cite{CC+T_1989} were carried out with the CFOUR 
suite of programs \cite{cfour,harding_JChemTheoryComput_4_64_2008}. 
The correlation consistent basis sets cc-pVXZ (X = T, Q), abbreviated XZ, were used for 
frozen core calculations \cite{cc-pVXZ_1989}, and weighted core-valence basis sets  were 
employed for calculation correlating all electrons (ae), these are the cc-pwCVXZ (X = T, Q, 5) 
basis sets, abbreviated wCXZ \cite{core-corr_2002}. All calculations were carried out at the 
Regionales Rechenzentrum der Universit{\"a}t zu K{\"o}ln (RRZK).

Equilibrium geometries were determined by analytic gradient techniques, harmonic 
force fields by analytic second derivatives, and anharmonic force fields by numerical 
differentiation of the analytically evaluated second derivatives of the energy. 
The main goal of these anharmonic force field calculations was to evaluate first order 
vibration-rotation parameters \cite{vib-rot_rev_1972}, see also Sect.~\ref{structure}. 
Core electrons were kept frozen unless ``ae'' indicates that all electrons were correlated.


\begin{table*}
\begin{center}
\caption{Ground state spectroscopic parameters$^{a}$ (MHz) of propyne isotopologs and of methyl cyanide.}
\label{ground-state-parameter}
\renewcommand{\arraystretch}{1.10}
\begin{tabular}[t]{lr@{}lr@{}lr@{}lr@{}lr@{}l}
\hline 
Parameter & \multicolumn{2}{c}{CH$_3$C$^{13}$CH} & \multicolumn{2}{c}{CH$_3^{13}$CCH} & \multicolumn{2}{c}{$^{13}$CH$_3$CCH} 
 & \multicolumn{2}{c}{CH$_3$CCH$^b$}  & \multicolumn{2}{c}{CH$_3$CN$^c$} \\
\hline
$(A - B)$                & 150850&.15          & 150598&.07          & 150827&.15          & 150594&.52~(46)    & 148900&.103~(66)    \\
$B$                      &   8290&.246538~(71) &   8542&.322556~(72) &   8313&.248883~(49) &   8545&.876998~(6) &   9198&.899167~(11) \\
$D_K \times 10^3$        &   2916&.            &   2916&.            &   2916&.            &   2915&.6~(122)    &   2830&.6~(18)      \\
$D_{JK} \times 10^3$     &    155&.49161~(282) &    162&.78978~(218) &    155&.21707~(197) &    163&.41706~(20) &    177&.40787~(25)  \\
$D_J \times 10^6$        &   2754&.725~(23)    &   2940&.771~(43)    &   2809&.959~(20)    &   2939&.320~(4)    &   3807&.576~(8)     \\
$H_K \times 10^6$        &    180&.            &    180&.            &    180&.            &    180&.           &    164&.6~(66)      \\
$H_{KJ} \times 10^6$     &      4&.9825~(202)  &      5&.2404~(126)  &      5&.0202~(194)  &      5&.2990~(17)  &      6&.0620~(14)   \\
$H_{JK} \times 10^9$     &    849&.74~(70)     &    911&.42~(111)    &    832&.13~(78)     &    915&.69~(13)    & 1\,025&.69~(15)     \\
$H_J \times 10^{12}$     &  $-$46&.7           &  $-$51&.2           &  $-$47&.1           &  $-$51&.2~(9)      & $-$237&.4~(21)      \\
$L_{KKJ} \times 10^{12}$ & $-$360&.            & $-$379&.            & $-$363&.            & $-$385&.3~(50)     & $-$444&.3~(25)      \\
$L_{JK} \times 10^{12}$  &  $-$40&.1           &  $-$43&.0           &  $-$41&.1           &  $-$43&.44~(47)    &  $-$52&.75~(51)     \\
$L_{JJK} \times 10^{12}$ &   $-$6&.07          &   $-$6&.65          &   $-$6&.12          &   $-$6&.648~(51)   &   $-$7&.901~(32)    \\
$L_J \times 10^{15}$     &       &             &       &             &       &             &       &            &   $-$3&.10~(17)     \\
$P_{JK} \times 10^{15}$  &       &             &       &             &       &             &       &            &      0&.552~(68)    \\
$P_{JJK} \times 10^{18}$ &     27&.            &     30&.            &     27&.            &     30&.4~(55)     &     55&.3~(22)      \\
\hline 
\end{tabular}\\[2pt]
\end{center}
$^a$ Numbers in parentheses are one standard deviation in units of the least significant figures. 
Parameters without quoted uncertainties have been estimated from the main isotopic species and were kept fixed in the fits; 
see section~\ref{Results}.\\
$^b$ The CH$_3$CCH parameters are also from this work; the value of $H_K$ was estimated from the CH$_3$CN value, the remaining values were obtained 
from a fit of previous data \cite{CH3CCH_rot_2000,CH3CCH_gs_vib_rot_IR_2002,MeC2_4H_rot_2008}; see also section~\ref{Results}.\\
$^c$ Ref.~\cite{MeCN_v8le2_2015}.
\end{table*}


\section{Observed spectra and determination of spectroscopic parameters}
\label{Results}

The $J = 7 - 6$ transitions of CH$_3$C$^{13}$CH and $^{13}$CH$_3$CCH were recorded near 
116~GHz. The corresponding transitions of CH$_3^{13}$CCH occur above the upper limit 
of the instrument at 118.2~GHz, and so the $J = 6 - 5$ transitions were measured instead 
for this isotopolog. For transitions with $0 \le K \le 3$, these transitions had been 
reported before \cite{propyne_wo_w-13C_rot_1978}, albeit with larger uncertainties of 
about 40~kHz, which were between 3 and 10~kHz in the present study. The quantum numbers 
of recorded transitions were also very similar for CH$_3$C$^{13}$CH and $^{13}$CH$_3$CCH 
in the submillimeter region and covered sufficiently strong transitions with 
$J = 26 - 25$, $35 - 34$, $45 - 44$, and $46 - 45$ between about 430~GHz and 763~GHz. 
The quantum numbers differed again somewhat for CH$_3^{13}$CCH; the transitions occured 
in sections between 426 and 785~GHz. The $K$ quantum numbers reached between 12 and 14, 
and the uncertainties were between 3 and 10~kHz for very good lines, up to 100~kHz for 
weaker or less symmetric lines. The experimental data employed in the fits are provided 
as supplementary material with quantum numbers, uncertainties and residuals between 
observed frequencies and those calculated from the final set of spectroscopic parameters.

Pickett's versatile programs SPFIT and SPCAT \cite{spfit_1991} were utilized for 
fitting and prediction of the rotational spectra. The purely axial parameters $A$ 
(or $A - B$), $D_K$, $H_K$, etc. of a prolate symmetric top can usually not be determined 
directly by rotational or rovibrational spectroscopy. One possibility to obtain these 
parameters involves $\Delta K = 3$ ground state loops generated from a doubly degenerate 
fundamental vibration and its combination and hot band associated with another doubly 
degenerate fundamental. The available data in the case of propyne were sufficient to 
determine $A$ and $D_K$ significantly \cite{CH3CCH_gs_vib_rot_IR_2002}. 
The similarity of the spectroscopic parameters of CH$_3$CCH with those of CH$_3$CN, 
in particular for the purely axial parameters, see Table~\ref{ground-state-parameter}, 
enabled us to estimate a value for $H_K$ of CH$_3$CCH. The value of $H_K$ of CH$_3$CN 
was scaled by the $D_K$ ratio of CH$_3$CCH and CH$_3$CN to the power of three which 
appeared to reflect better the trends observed in higher order parameters than scaling 
by that ratio to the power of three half. The remaining parameters were redetermined 
from earlier data \cite{CH3CCH_rot_2000,CH3CCH_gs_vib_rot_IR_2002,MeC2_4H_rot_2008}, 
but differ only slightly from those of an earlier combined fit \cite{MeC2_4H_rot_2008}.


\begin{table*}
\begin{center}
\caption{Ground state rotational parameters $B_0^i$ (MHz) of propyne isotopologs, semi-empirical equilibrium 
         rotational parameters $B_e^i$ (MHz) according to two quantum-chemical models, residuals O$-$C (MHz) 
         between $B_e^i$ and values from the structure calculations, and references$^a$ Refs. from which the 
         $B_0^i$ were derived.}
\label{r_e-emp-parameter}
\renewcommand{\arraystretch}{1.10}
\begin{tabular}[t]{lcr@{}lr@{}lr@{}lcr@{}lr@{}ll}
\hline
 & & & & \multicolumn{4}{c}{CCSD(T)/cc-pVTZ} & & \multicolumn{4}{c}{CCSD(T)/cc-pwCVTZ} & \\ 
\cline{5-8} \cline{10-13}
Isotopolog & Axis $i$ & \multicolumn{2}{c}{$B_0^i$} & \multicolumn{2}{c}{$B_e^i$} & 
\multicolumn{2}{c}{O$-$C} & & \multicolumn{2}{c}{$B_e^i$} & \multicolumn{2}{c}{O$-$C} & Refs. \\
\hline
CH$_3$CCH         & $b$  & 8545&.877 & 8573&.176 & $-$0&.0286 & & 8573&.231 & $-$0.&0316 & TW \\
CH$_3$C$^{13}$CH  & $b$  & 8290&.247 & 8316&.680 & $-$0&.0197 & & 8316&.741 & $-$0.&0215 & TW \\
CH$_3^{13}$CCH    & $b$  & 8542&.333 & 8569&.651 &    0&.0099 & & 8569&.707 &    0.&0083 & TW \\
$^{13}$CH$_3$CCH  & $b$  & 8313&.249 & 8338&.960 & $-$0&.0248 & & 8339&.016 & $-$0.&0278 & TW \\
CH$_3$CCD         & $b$  & 7788&.169 & 7808&.965 & $-$0&.0065 & & 7808&.999 & $-$0.&0070 & \cite{CH2DCCH_CH3CCD_isos_rot_1993}; \cite{propyne_div-ios_rot_1950,MeCCH_MeCN_div-isos_rot_1955,CH3CCD_CD3CCH_rot_HFS_1982} \\
CH$_3$C$^{13}$CD  & $b$  & 7592&.916 & 7613&.230 &    0&.0021 & & 7613&.267 & $-$0.&0004 & \cite{CH2DCCH_CH3CCD_isos_rot_1993} \\
CH$_3^{13}$CCD    & $b$  & 7787&.026 & 7807&.844 &    0&.0260 & & 7807&.879 &    0.&0267 & \cite{CH2DCCH_CH3CCD_isos_rot_1993} \\
$^{13}$CH$_3$CCD  & $b$  & 7576&.800 & 7596&.329 & $-$0&.0023 & & 7596&.364 & $-$0.&0033 & \cite{CH2DCCH_CH3CCD_isos_rot_1993} \\
CH$_2$DCCH        & $b$  & 8155&.686 & 8180&.361 &    0&.0027 & & 8180&.412 &    0.&0053 & \cite{CH2DCCH_CH3CCD_isos_rot_1993}; \cite{MeCCH_MeCN_div-isos_rot_1955,CH2DCCH_rot_1990} \\
                  & $c$  & 8025&.476 & 8053&.941 & $-$0&.0011 & & 8053&.949 & $-$0.&0011 &  \\
CH$_2$DC$^{13}$CH & $b$  & 7908&.447 & 7932&.379 &    0&.0122 & & 7932&.435 &    0.&0151 & \cite{CH2DCCH_CH3CCD_isos_rot_1993} \\
                  & $c$  & 7785&.911 & 7813&.422 &    0&.0079 & & 7813&.437 &    0.&0081 &  \\
CH$_2$D$^{13}$CCH & $b$  & 8150&.513 & 8175&.206 &    0&.0408 & & 8175&.258 &    0.&0450 & \cite{CH2DCCH_CH3CCD_isos_rot_1993} \\
                  & $c$  & 8020&.426 & 8048&.895 &    0&.0327 & & 8048&.904 &    0.&0343 &  \\
$^{13}$CH$_2$DCCH & $b$  & 7956&.786 & 7980&.122 &    0&.0026 & & 7980&.174 &    0.&0047 & \cite{CH2DCCH_CH3CCD_isos_rot_1993} \\
                  & $c$  & 7832&.349 & 7859&.325 & $-$0&.0014 & & 7859&.335 & $-$0.&0025 &  \\
CH$_2$DCCD        & $b$  & 7440&.765 & 7459&.582 &    0&.0041 & & 7459&.613 &    0.&0075 & \cite{MeCCH_MeCN_div-isos_rot_1955} \\
                  & $c$  & 7331&.974 & 7354&.098 & $-$0&.0056 & & 7354&.093 & $-$0.&0049 & \\
CHD$_2$CCH        & $b$  & 7765&.701 & 7789&.873 & $-$0&.0264 & & 7789&.896 & $-$0.&0288 & \cite{MeCCH_MeCN_div-isos_rot_1955} \\
                  & $c$  & 7630&.949 & 7658&.396 & $-$0&.0089 & & 7658&.383 & $-$0.&0056 & \\
CD$_3$CCH         & $b$  & 7355&.701 & 7380&.982 &    0&.0004 & & 7380&.963 & $-$0.&0035 & \cite{MeCCH_MeCN_div-isos_rot_1955}; \cite{CD3CCH_etc_rot_dip_1970,CH3CCD_CD3CCH_rot_HFS_1982} \\
CHD$_2$CCD        & $b$  & 7095&.074 & 7113&.693 & $-$0&.0152 & & 7113&.701 & $-$0.&0165 & \cite{MeCCH_MeCN_div-isos_rot_1955} \\
                  & $c$  & 6982&.554 & 7004&.047 & $-$0&.0059 & & 7004&.026 & $-$0.&0018 & \\
CD$_3$CCD         & $b$  & 6734&.340 & 6754&.068 &    0&.0065 & & 6754&.041 &    0.&0030 & \cite{MeCCH_MeCN_div-isos_rot_1955}; \cite{propyne_div-ios_rot_1950} \\
\hline 
\end{tabular}\\[2pt]
\end{center}
$^a$ TW stands for this work. The primary source is given for data not from this work. 
     Additional references, as far as applicable, are separated from the primary one by a semi-colon.
\end{table*}


The purely axial parameters $A$, $D_K$, and $H_K$ of the propyne isotopomers with one $^{13}$C 
were fixed to values of the main isotopolog because substitution of an atom on the symmetry axis 
does not change $A_e$, and potential changes in $D_K$ and $H_K$ should be very small. 
Several higher order parameters were scaled to appropriate powers of the isotopic ratios 
of the $B$ values in a first step. Trends in the small deviations observed for quartic and 
some sextic distortion parameters were taken into account in a second step. 
Such empirically scaled parameters are usually much better than fixing these higher order 
parameters to zero and often also better than fixing the values to those of the main isotopic 
species, as also shown for isotopic species of methyl cyanide \cite{MeCN_isos_v8_rot_2016}. 
The relations hold strictly for the Dunham parameters of a diatomic molecule. 
The empirical scaling is more complex for asymmetric rotors of the prolate type for which 
appropriate powers of $A - (B + C)/2$, $B + C$ and $B - C$ would be used, see the examples 
of vinyl cyanide \cite{13C-VyCN_2008} or thioformaldehyde \cite{H2CS_rot_2019}. 
The resulting spectroscopic parameters of the singly $^{13}$C substituted propyne isotopomers 
are given in Table~\ref{ground-state-parameter} together with those of CH$_3$CCH and CH$_3$CN. 
The uncertainties of a particular parameter are of the same order of magnitude among the 
three $^{13}$C species, and differences reflect variations in the data sets. As the propyne 
and methyl cyanide molecules are isoelectronic, they share very similar structures. 
The $A$ rotational parameters are very similar, 159.1~GHz versus 158.1~GHz, and so are 
the $D_K$ parameters. The structural changes caused by the CH group in propyne versus 
the N atom in methyl cyanide can account for the slightly larger differences in the $B$ 
rotational parameters, and the differences increase further with increasing $J$ dependence 
(increasing power of $J(J+1)$) of a given parameter.

\section{Structural parameters of propyne}
\label{structure}

The equilibrium structure is the best and easiest defined structure of a molecule. 
It requires to calculate equilibrium rotational parameter(s), for example, $B_e$ 
from the ground state rotational parameter(s) $B_0$ as follows

\begin{equation}
\label{equi-B}
B_e = B_0 + \frac{1}{2}\sum_{j} \alpha _j^B - \frac{1}{4}\sum_{j \le k} \gamma _{jk}^B - ...
\end{equation}

\noindent
where the $\alpha _j^B$ are first order vibration-rotation interaction parameters, the 
$\gamma _{jk}^B$ are second order vibration-rotation interaction parameters, and so on. 
Doubly degenerate modes occur two times in the sum. Equivalent formulations hold for $A_e$ 
(and $C_e$ in the case of asymmetric rotor isotopic species). The general $n$-atomic asymmetric 
rotor molecule has three different rotational parameters $A$, $B$, and $C$, $3n - 6$ first order 
vibrational corrections, $(3n - 6)(3n - 5)/2$ second order vibrational corrections, and so on. 
The situation is slightly better for symmetric top molecules. In addition, data for more than 
one isotopic species need to be known to determine all independent structural parameters, 
unless the molecule is a symmetric triatomic of the type AB$_2$, where atoms A and B may be 
the same. Complete sets even of first order vibration-rotation parameters for a sufficiently 
large ensemble of isotopic species are available from experiment only in rare cases. Moreover, 
these parameters are not identical to the $\alpha _j^B$ because of effects of the $\gamma _{jk}^B$ 
etc. and possibly because of vibration-rotation interactions such as Fermi or Coriolis resonances.


\begin{table*}
\begin{center}
\caption{Quantum-chemical and experimental bond lengths (pm) and bond angle (deg) of propyne.$^a$}
\label{struct-parameters}
\renewcommand{\arraystretch}{1.10}
\begin{tabular}[t]{lr@{}lr@{}lr@{}lr@{}lr@{}l}
\hline
Method$^b$ & \multicolumn{2}{c}{$r$(C$_{\rm m}$H)} & \multicolumn{2}{c}{$r$(C$-$C)} & \multicolumn{2}{c}{$\angle$(HCC)} & 
  \multicolumn{2}{c}{$r$(C$\equiv$C)} & \multicolumn{2}{c}{$r$(C$_{\rm a}$H)} \\
\hline
CCSD(T)/TZ                 & 109&.087     & 146&.653     & 110&.577     & 121&.114     & 106&.276     \\
CCSD(T)/QZ                 & 108&.979     & 146&.366     & 110&.580     & 120&.810     & 106&.290     \\
ae-CCSD(T)/wCTZ            & 108&.908     & 146&.242     & 110&.602     & 120&.706     & 106&.244     \\
ae-CCSD(T)/wCQZ            & 108&.824     & 146&.003     & 110&.611     & 120&.504     & 106&.156     \\
ae-CCSD(T)/wC5Z            & 108&.790     & 145&.935     & 110&.602     & 120&.450     & 106&.128     \\
CEPA-1/TZ$^c$              & 108&.77      & 147&.06      & 110&.50      & 120&.65      & 106&.17      \\
dito, refined$^d$          & 108&.77      & 145&.84      & 110&.50      & 120&.53      & 106&.17      \\
$r_0$$^e$                  & 109&.40~(4)  & 145&.95~(5)  & 110&.6~(2)   & 120&.88~(6)  & 105&.48~(3)  \\
$r_{I,\epsilon}$$^e$       & 109&.27~(3)  & 145&.74~(1)  & 110&.8~(1)   & 120&.75~(1)  & 105&.62~(1)  \\
$r_m^{\rho}$$^e$           & 108&.85~(5)  & 145&.52~(5)  & 111&.12~(6)  & 120&.37~(6)  & 105&.89~(3)  \\
$r_e^{\rm SE}$(CCSD(T))    & 108&.836~(3) & 145&.884~(6) & 110&.593~(2) & 120&.463~(6) & 106&.141~(3) \\
$r_e^{\rm SE}$(ae-CCSD(T)) & 108&.852~(4) & 145&.884~(6) & 110&.591~(2) & 120&.460~(7) & 106&.146~(3) \\
\hline
\end{tabular}\\[2pt]
\end{center}
$^a$ All values from this work unless indicated otherwise. Numbers in parentheses are one standard deviation 
     in units of the least significant figures. The methylenic and acetylenic CH bonds are indicated by 
     C$_{\rm m}$H and C$_{\rm a}$H, respectively.\\
$^b$ Quantum-chemical calculations as detailed in Sect.~\ref{qcc}.\\ 
$^c$ Coupled electron pair approximation \cite{CEPA_1976}, data from Ref.~\cite{propyne_ai_1996}.\\
$^d$ CEPA-1/TZ, refined; CC bonds corrected empirically \cite{propyne_ai_1996}.\\
$^e$ Ref.~\cite{CH2DCCH_CH3CCD_isos_rot_1993}; see section~\ref{structure} for explanation on the 
     structure models.
\end{table*}


\begin{table*}
\begin{center}
\caption{Equilibrium bond lengths (pm) and bond angle (deg) of propyne in comparison to 
         values of related molecules.$^a$}
\label{comp-struct}
\renewcommand{\arraystretch}{1.10}
\begin{tabular}[t]{lr@{}lr@{}lr@{}lr@{}lr@{}l}
\hline
Molecule    & \multicolumn{2}{c}{$r$(C$_{\rm m}$H)} & \multicolumn{2}{c}{$r$(C$-$C)} & \multicolumn{2}{c}{$\angle$(HCC)} & 
  \multicolumn{2}{c}{$r$(C$\equiv$C)} & \multicolumn{2}{c}{$r$(C$_{\rm a}$H)} \\
\hline
CH$_3$CCH$^b$     & 108&.852~(4) & 145&.884~(6) & 110&.591~(2) & 120&.460~(7)   & 106&.146~(3)  \\
CH$_3$CN$^c$      & 108&.65~(1)  & 145&.85~(4)  & 109&.84~(1)  &    &           &    &          \\
CH$_3$C$_4$H$^d$  & 108&.90~(24) & 145&.69~(8)  & 110&.50~(8)  & 120&.85~(6)    & 106&.13~(3)   \\
C$_2$H$_3$CCH$^e$ &    &         & 142&.67      &    &         & 120&.72        & 106&.17       \\
HC$_4$H$^f$       &    &         & 137&.27      &    &         & 120&.85        & 106&.15       \\
HCCH$^g$          &    &         &    &         &    &         & 120&.2958~(7)  & 106&.164~(1)  \\
HCCH$^h$          &    &         &    &         &    &         & 120&.2817~(12) & 106&.167~(14) \\
CH$_3$F$^i$       & 108&.70~(5)  &    &         &    &         &    &           &    &          \\
C$_2$H$_6$$^j$    & 108&.9~(1)   & 152&.2~(2)   & 111&.2~(1)   &    &           &    &          \\
\hline
\end{tabular}\\[2pt]
\end{center}
$^a$ Numbers in parentheses are one standard deviation in units of the least significant figures, 
     unless stated otherwise. The methylenic and acetylenic CH bonds are indicated by 
     C$_{\rm m}$H and C$_{\rm a}$H, respectively.\\
$^b$ This work; $r_e^{\rm SE}$, ae-CCSD(T)/wCTZ values.\\ 
$^c$ Ref.~\cite{CH3CN_re-SE_2006}; $r_e^{\rm SE}$, ae-CCSD(T)/wCTZ values with 3$\sigma$ uncertainties.\\
$^d$ Ref.~\cite{CH3C4H_re-SE_2008}; $r_e^{\rm SE}$, MP2/TZ values with 3$\sigma$ uncertainties; 
     see also discussion in section~\ref{structure}.\\
$^e$ Ref.~\cite{VyX_struktur_2018}; $r_e^{\rm SE}$, CCSD(T)/TZ values with no uncertainties given.\\
$^f$ Ref.~\cite{C4H2_re-SE_2008}; $r_e^{\rm SE}$, ae-CCSD(T)/CQZ values with no uncertainties given.\\
$^g$ Ref.~\cite{C2H2_re-SE_2011}; $r_e^{\rm SE}$, ae-CCSD(T)/wCQZ values.\\
$^h$ Ref.~\cite{C2H2_re-exp_2016}; experimental $r_e$ values.\\
$^i$ Ref.~\cite{re_CH3F_1999}; experimental $r_e$ value.\\
$^j$ Ref.~\cite{ethane_r_1990}; $r_m^{\rho}$ values according to the second set of data.
\end{table*}


An alternative, lately very common, approach is to calculate $\sum_{j} \alpha _j^B$ by 
quantum-chemical means to derive semi-empirical equilibrium rotational parameters $B_{i,e}$ 
from the experimental ground state values \cite{r_e_emp_1998,vazquez_equilibrium_structures_2011}. 
Second and higher order vibrational contributions are neglected. Numerous quantum-chemical 
programs are available to carry out such calculations.

We evaluated $\sum_{j} \alpha _j^B$ employing CCSD(T)/TZ and ae-CCSD(T)/wCTZ calculations for all 
available isotopologs; the corresponding data are summarized in Table~\ref{r_e-emp-parameter}. 
The ground state rotational parameters, the vibrational corrections calculated at two different 
levels, and the resulting semi-empirical equilibrium rotational parameters are available as 
supplementary material. 
The rotational parameters obtained in earlier studies were refit in the present study taking into 
account estimates of (higher order) centrifugal distortion parameters in order to reduce effects 
from omission or truncation of these parameters. Furthermore, we used the $S$ reduction for all 
asymmetric rotor isotopologs.

Structural parameter were determined using the program STRFIT \cite{strfit_2003}. 
The semi-empirical equilibrium moments of inertia derived from the respective rotational 
parameters in Table~\ref{r_e-emp-parameter} were adjusted. Data pertainting to the $a$-axis 
were omitted because of lack of accuracy. The resulting structural parameters $r_e^{\rm SE}$ 
are given in Table~\ref{struct-parameters} together with values from quantum-chemical 
calculations and from previous structure determinations. The $r_e^{\rm SE}$ equilibrium structural 
parameters of propyne are compared with values of related molecules in Table~\ref{comp-struct}.

The CCSD(T)/XZ bond lengths are longer than the ae-CCSD(T)/wCXZ values, and both display 
the usual bond shortening upon increase in basis set size. The bond angles are quite similar 
throughout. The ae-CCSD(T)/wC5Z structural parameters are very close to our semi-empirical 
structures, and both models yield virtually the same values with the exception of the 
methylenic CH bond, for which the values differ by slightly more than two times the combined 
uncertainties. The CEPA-1/TZ structural parameters agree quite well with our $r_e^{\rm SE}$ 
values, except for the CC single bond. CEPA stands for coupled electron pair approximation. 
It is a quantum-chemical method described in Ref.~\cite{CEPA_1976}. The agreement is better 
after empirical refining of the CC bond lengths. The correction was based on a comparison 
between CEPA-1/TZ and experimental CC bond lengths of diacetylene \cite{propyne_ai_1996}. 
The ground state effective $r_0$ structure agrees modestly well with our $r_e^{\rm SE}$ values. 
The agreement improves somewhat for the $r_{I,\epsilon}$ values and more so for the 
$r_m^{\rho}$ values. The $r_{I,\epsilon}$ structure model assumes $I_{i,0} - I_{i,e}$ to be 
isotopic independent for a given axis $i$; it is equivalent to the substitution structure 
$r_s$, as investigated in Ref.~\cite{r_I-eps_1991}. The $r_m^{\rho}$ structure model assumes 
a more complex isotopic scaling of these differences, see Ref.~\cite{r-rho-m_1986} for more 
details.

The CC triple bond in acetylene has a length of 120.29~pm \cite{C2H2_re-SE_2011,C2H2_re-exp_2016}; 
delocalization of $\pi$ electrons between the two CC triple bonds and the CC single bond in 
diacetylene lengthens the triple bonds to 120.85~pm and shortens the single bond considerably 
to 137.27~pm \cite{C4H2_re-SE_2008} from 152.2~pm in ethane \cite{ethane_r_1990}. Only the methyl 
group in propyne can be involved in the delocalization of the $\pi$ electrons of the CC triple bond, 
which leads to a triple bond length of 120.46~pm, between the acetylene and diacetylene values, and 
to a single bond length of 145.88~pm, between the diacetylene and the ethane values. 
The degree of delocalization is larger in vinyl acetylene than in propyne, leading to CC triple 
and single bond lengths closer to those of diacetylene \cite{VyX_struktur_2018}. The acetylenic 
CH bond lengths in Table~\ref{comp-struct} are all very close to 106.15~pm; slightly larger 
scatter exists for the given methylenic CH bond lengths. The structural parameters of the CH$_3$C 
unit in methyl cyanide \cite{CH3CN_re-SE_2006} are quite similar to those of the isoelectronic 
propyne. In addition, the parameters of the CH$_3$C unit in pentadiyne \cite{CH3C4H_re-SE_2008} 
and propyne are quite similar also. The CC triple bond lengths of $120.91 \pm 0.16$~pm and 
$120.85 \pm 0.06$~pm and the length of the CC single bond between them ($137.34 \pm 0.14$~pm) 
are very close to the corresponding diacetylene values, as may be expected.


\begin{table}
\begin{center}
\caption{Predicted main ground state spectroscopic parameters$^a$ (MHz) of propyne isotopomers with two $^{13}$C.}
\label{2x13-parameter}
\renewcommand{\arraystretch}{1.10}
\begin{tabular}[t]{lr@{}lr@{}lr@{}l}
\hline 
Parameter & \multicolumn{2}{c}{CH$_3^{13}$C$^{13}$CH} & \multicolumn{2}{c}{$^{13}$CH$_3$C$^{13}$CH} & \multicolumn{2}{c}{$^{13}$CH$_3^{13}$CCH} \\
\hline
$B^b$                    &   8288&.214 &   8060&.023 &   8308&.459 \\
$B^c$                    &   8288&.234 &   8060&.007 &   8308&.472 \\
$D_{JK} \times 10^3$     &    155&.04  &    147&.53  &    154&.50  \\
$D_J \times 10^6$        &   2755&.4   &   2630&.3   &   2811&.7   \\
$H_{KJ} \times 10^6$     &      4&.93  &      4&.72  &      4&.96  \\
$H_{JK} \times 10^9$     &    849&.    &    911&.    &    832&.    \\
\hline 
\end{tabular}\\[2pt]
\end{center}
$^a$ See section~\ref{2x13C-propyne}.\\
$^b$ Calculated $B_e$ from the semi-empirical CCSD(T)/cc-pwCVTZ structure and vibrational corrections 
     from the CCSD(T)/cc-pwCVTZ calculation.\\
$^c$ As above, but also corrected for the residual of the CH$_3$CCH isotopolog and for the differences between 
     the main isotopolog and the appropriate $^{13}$C isotopomers.
\end{table}

\section{Spectroscopic parameters for propyne isotopomers with two $^{13}$C atoms}
\label{2x13C-propyne}

We evaluated spectroscopic parameters for propyne isotopomers containing two $^{13}$C, 
because their rotational spectra have not yet been investigated, but they may be detectable 
in astronomical sources with very high enrichment in $^{13}$C compared to the dominant 
$^{12}$C, such as in the envelope of the planetary nebula K4$-$47 in which the 
$^{12}$C/$^{13}$C ratio is as low as $2.2 \pm 0.8$ and in which all isotopomers of 
propyne with one $^{13}$C were already detected \cite{PN_K4-47_2019}.

In a first step, the semi-empirical rotational parameter $B_e$ was calculated for each of 
the three isotopomers from the ae-CCSD(T)/wCTZ semi-empirical structure. The first order 
vibrational corrections calculated at the same level were subtracted off to yield the 
first estimate of the ground state rotational parameter of each of the three isotopomers 
given in the first row of Table~\ref{2x13-parameter}. The semi-empirical rotational parameters 
of the various isotopologs of propyne were not fit perfectly well, but have small residuals on 
the order of a few tens of kilohertz, as can be seen in Table~\ref{r_e-emp-parameter}. 
We sought to improve the estimates of the ground state rotational parameters of the three 
isotopomers with two $^{13}$C atoms by correcting each $B$ value with the residual of the 
main isotopolog and then by the differences in the residuals between the appropriate 
isotopologs with one $^{13}$C and the main species. This yields the $B$ values given 
in the second row of Table~\ref{2x13-parameter}.

The quartic centrifugal distortion parameters $D_{JK}$ and $D_{J}$ as well as the sextic 
parameters $H_{KJ}$ and $H_{JK}$ were evaluated in a fairly similar way. The quantum-chemically 
calculated equilibrium quartic centrifugal distortion parameters for isotopomers with one and 
two $^{13}$C atoms were corrected by the ratio of the experimental ground state $D$ and the 
calculated equilibrium $D$. This yields good estimates for the experimental ground state $D$ 
values of the $^{13}$C containing isotopologs. The values of the doubly substituted isotopic 
species were then corrected with the ratios between the experimentally determined $D$ values 
of the appropriate singly substituted isotopologs and their first order estimates determined 
in the previous step. These values should be very good estimates of $D_{JK}$ and $D_{J}$ for 
the propyne isotopomers with two $^{13}$C. The sextic centrifugal distortion parameters $H_{KJ}$ 
and $H_{JK}$ were evaluated from the experimental values of the main isotopic species by applying 
the ratios between the appropriate singly substituted species and the main species. The resulting 
distortion parameters are also given in Table~\ref{2x13-parameter}.

\section{Conclusions}
\label{Conclusions}

Our measurements of the rotational spectra of propyne isotopomers containing one $^{13}$C 
increased the upper limit of data from $\sim$240~GHz to $\sim$785~GHz, yielding greatly 
improved spectroscopic parameters which should allow reasonable extrapolation to about 
1.5~THz, beyond the current upper limit of ALMA and possibly well beyond the needs of 
radio astronomers. This estimate is based on observations that in $a$-type spectra with 
good data coverage at the upper frequency limit the calculated uncertainties are still 
meaningful at about two times the upper frequency limit. The calculated uncertainties at 
1.5~THz are $\sim$60~kHz or less at low $K$ up to $\sim$200~kHz at $K = 10$. 
Thus, we expect that actual transition frequencies will be within $\sim$1~MHz of 
the calculations at this frequency.
The line, parameter and fit files are available in the data 
section\footnote{See: https://cdms.astro.uni-koeln.de/classic/predictions/daten/CH3CCH/} 
of the CDMS\cite{CDMS_2016}; calculations of the rotational spectra are deposited in its 
catalog section\footnote{See: https://cdms.astro.uni-koeln.de/classic/entries/}.
Quantum-chemical calculations yielded first order vibration-rotation interaction parameters, 
which were used to derive semi-empirical equilibrium structural parameters of propyne. 
These parameters are in very good agreement with quantum-chemical calculations at 
the highest level employed in the present study. Finally, we estimated the main 
spectroscopic parameters of propyne isotopomers containing two $^{13}$C to facilitate 
their detection in astronomical source with high levels of $^{13}$C or to identify them 
in laboratory measurements.


\section*{Acknowledgments}

We are grateful for support by the Deutsche Forschungsgemeinschaft via the collaborative 
research centers SFB~494, subproject E2, and SFB~956 (project ID 184018867), project B3 
and via the Ger{\"a}tezentrum SCHL~341/15-1 (``Cologne Center for Terahertz Spectroscopy''). 
We thank the Regionales Rechenzentrum der Universit{\"a}t zu K{\"o}ln (RRZK) for providing 
computing time on the DFG funded High Performance Computing System CHEOPS. 
Our research benefited from NASA's Astrophysics Data System (ADS).

\appendix
\section*{Appendix A. Supplementary Material}

The following are the Supplementary data to this article: 
The part of the fit files which contain all experimental data employed in the fits 
with quantum numbers, uncertainties and residuals between observed frequencies and 
those calculated from the final sets of spectroscopic parameters. Also included in 
these files are the spectroscopic parameters with uncertainties and correlation 
coefficients. A separate file gives the ground state rotational parameters, 
the vibrational corrections calculated at two different levels, and the resulting 
semi-empirical equilibrium rotational parameters of all isotopic species used for 
the structure evaluations.



\bibliographystyle{elsarticle-num}
\bibliography{propin}






\end{document}